Emergent Explicit Regulation in College Students' Collaborative Scientific Inquiry Learning—Framework and A Case Study

Ying Cao[1*], Tong Wan[2], Andrew Burns[3], Ethan Cusack[4], and Pierre-Philippe A. Ouimet[5]

1. School of Education and Child Development, Drury University, 900 N. Benton Ave. Springfield, MO 65802, USA
2. Department of Physics, University of Central Florida, 4111 Libra Drive, Orlando, FL 32816, USA
3. Department of Chemistry and Physics, Drury University, 900 N. Benton Ave. Springfield, MO 65802, USA
4. Department of Language and Literature, Drury University, 900 N. Benton Ave. Springfield, MO 65802, USA
5. Department of Physics, University of Regina, 3737 Wascana Parkway, Regina, Saskatchewan, S4S 0A2, Canada

Abstract
Small-group activities have been widely adopted in college level science courses. As students participate in these activities, it is important to consider how group members collectively regulate their activity and complete group task. Regulation in a group often involves adaptive responsivity from group members when they notice and deal with a challenge. The theoretical framework of socially shared regulation emphasizes group members collaboratively regulating within the group but does not focus on portraying how the shared regulation is developed in the moment. Currently, the field lacks a framework characterizing the momentary development of a regulatory action in a group. In our study, we are trying to address this gap. In our video data, incoming first-year college students were enrolled in a summer program designed to promote students' metacognitive skills to be incorporated in their study of science. The students were first generation and/or deaf and hard of hearing (DHH), and were experiencing small-group, inquiry-based, hands-on science activities. We have observed various moments in which the students spontaneously made a move to regulate (i.e., to direct or to adjust) the activity in completing their tasks and achieving group goals. We subsequently have developed a framework called Emergent Explicit Regulation, or EER, to characterize those moments. The EER framework aligns with the literature about socially shared regulative learning; it captures students' in-the-moment regulatory moves to respond to a challenge—articulating how those moves emerge, in what ways they are explicit and regulatory. In this paper, we first introduce the EER framework that we developed through our iterative analyses of the large pool of the program video data, and situate the EER framework in the context of collaborative scientific inquiry learning. We then present a case study where we applied the EER framework to identify typical EER instances in one small group when the students completed the task of building a model to represent the climate of the Earth's atmosphere. The students in this group worked collaboratively, faced and handled various challenges, and completed the group task. They demonstrated multiple EERs in different psychological areas and in the inquiry practices designed in the activity. Our case study demonstrates how the EER framework can be used to examine the regulatory moves made by group members toward achieving shared learning goals. Tentative instructional implications are suggested based on the findings.

*Corresponding author: ycao@drury.edu

## Introduction

Small group activities play an important role in science education. Such activities are key in a wide range of pedagogical approaches, from traditional labs to modern studio classrooms (e.g., Brookes *et al.* 2021; Stewart *et al.*, 2016; & Van De Bogart *et al.*, 2017). In observing groups engaged in science learning activities we have noticed that some are more productive than others in terms of completing group tasks and meeting group goals, even when these different groups are in the same program, taught by the same instructor, and are working on the same tasks. One approach to unpacking group activities and learning more about what makes a group more productive is through examining students' regulations in their group (Hadwin *et al.*, 2017; Sobocinski *et al.*, 2020 & Sobocinski *et al.*, 2022). Regulation in learning, as one demonstration of metacognition (Brown, 1987), is the process where learners orient themselves towards the attainment of learning goals. In inquiry-based science activities, the manner in which groups regulate their activity is especially crucial due to the open-endedness of the tasks. Students must work collectively, monitor and adjust when needed in order to achieve shared goals.

Our research study was based on a rich set of video data, generated in the IMPRESS (Integrating Metacognitive Practices and Research to Ensure Student Success) summer program. The IMPRESS program is a two-week program at Rochester Institute of Technology (RIT) for matriculating students who are first generation and/or deaf and hard of hearing (DHH). It is designed to serve as a bridge program for students to learn how to reflect on, evaluate, and change their own thinking through intensive laboratory experiments, reflective practices, and discussion both in small groups and with the whole class. The main objectives of the IMPRESS program are to engage students in authentic, inquiry-based scientific practice, to facilitate the development of a supportive community, and to help the students reflect on science and themselves to strengthen their learning habits and lead them to a successful career in STEM fields. Four years of classroom video data were collected through IMPRESS, from which we extracted and studied instances of student groups' regulatory behavior in science activities.

In our analysis of the IMPRESS data, we noticed several instances of students' spontaneous regulation in their activities. We examined these cases and name the phenomenon *emergent explicit regulation*, or EER (Cao *et al.*, 2019 & Ouimet *et al.*, 2022). We defined EER as student-made *regulatory moves*—meaning students take actions to direct or adjust the activity to handle a challenge or concern arises from their evaluation of the situation at the moment. An EER is *emergent*—meaning it is students' in-the-moment, self-initiated responses to a challenge or concern (in contrast to instructed or previously planned regulations); it is *explicit*—meaning students articulate their ideas and/or take observable actions to direct or adjust the activity.

EER captures the moments in which a regulatory response is developed from implicit to explicit—i.e., from where group members implicitly monitor and evaluate the situation, recognize a challenge or concern that needs to be addressed, to where the group explicitly take actions directing or adjusting the activity. Instead of using the term regulation in a broader sense, which can include planning, monitoring, and evaluating (Tanner, 2012), we use regulation in the EER framework narrowly to mean this adaptive response. The challenge or concern can be the group finding it hard to get started, group members having different ideas but needing to pick one and move forward, the group not knowing how to make sense of the data, not knowing how to use or adjust equipment, or just experiencing an awkward silence. Facing a group challenge or a concern, an individual either directly regulates (e.g., asks the group to do something or to stop doing something) or explicitly expresses their concern (e.g., asks a question) that may lead to a regulation. The individual or individuals demonstrate metacognitive skills in monitoring the group task and bringing up a potential change in the current flow of the activity, intending to better complete the group task. Multiple instances of such regulation can show patterns of regulation—for example, these instances are distributed among different group members when the group works collaboratively or concentrated in a small subset of the group members when there is dominance in regulating the group. Understanding the pattern of EER distribution can inform design and intervention aimed at fostering effective and collaborative self-regulated learning.

Grounded on ample instances of EER we collected over these years of study (Cao *et al.*, 2019 & Ouimet *et al.*, 2022), we subsequently formalized EER into a theoretical framework to analyze more emergent regulations in different contexts. EER fits in socially shared regulation—i.e., when group members collaboratively regulate within the group (Hardwin *et al.*, 2017)—and primarily focuses on its demonstration in the contexts of science learning. Vauras *et al.* (2021) argue that science is a highly challenging discipline which offers an intriguing window to unveil metacognitive processes emerging in social contexts. Vauras *et al.* called for more empirically grounded evidence to fully understand socially shared regulation as a joint effort of all the participants to reach the intended goals in science learning.

In this paper, we first describe the EER framework and show how EER can be applied to analyze the momentary regulations in a collaborative, inquiry-oriented science learning context. When describing the framework, we draw examples from the IMPRESS video data to help us articulate the framework.

We then present a case study applying EER to analyze a small group in two consecutive days where the students engaged in a part of a sequence of hands-on, inquiry-based science activities—building a physical model to represent Earth's atmosphere and the greenhouse Effect. The IMPRESS program has generated several hundred hours of video data that we can use for our analysis. Upon a quick scan of the IMPRESS data, we found that this group worked more collaboratively, faced and handled various challenges, and demonstrated metacognitive and regulative skills. Their frequent verbalization and exchange of ideas allowed us to identify their explicit regulations more accurately. This group had two students self-identified as members of DHH community, and one of these two signed to communicate with the rest of the group via an interpreter. DHH students and signing communication were featured in the IMPRESS data, and this group reflected that feature. Findings yielded from a group of mixed communication methods can have implications for collaborative groups where members speak different languages to communicate, a feature of many real-world scientific collaborations. In the case study, we ask the following research question: in a small group of students using mixed communication methods engaging in hands-on, inquiry-oriented science activities, in what ways do the students regulate their activities via emergent explicit regulation?

## Literature Review

*Metacognition and Self-Regulation*

Our study about regulation in a group falls into the area of research concerning student metacognition. Metacognition can be referred to as one's "knowledge concerning one's own cognitive processes or anything related to them" (Flavell, 1979). It has been interpreted as a combination of knowledge of cognition, and regulation of cognition (Brown, 1987) and applied to analyze undergraduate students' metacognitive skills when they completed homework in a team-based learning environment (Mota *et al*., 2019). When students monitor and direct their own learning progress, they ask questions such as "What am I doing now?" "Is it getting me anywhere?" "What else could I be doing instead?" (Perkins & Salomon, 1989). These questions are references to the metacognitive self. Students' spontaneous metacognitive talks (Sayre & Irving, 2015) and self-efficacy (Quan & Elby, 2016) have been studied in physics education research.

Self-regulation in cognition and learning—one demonstration of metacognitive skills—is defined as "the ability to orchestrate one's learning: to plan, monitor success, and correct errors when appropriate–all necessary for effective intentional learning" (NRC, 2000). The process of planning, monitoring, and evaluating has been studied in multiple areas such as students' mathematical problem solving in (Polanyi, 1957; Gray, 1991), mathematical thinking skills (Tanner & Jones, 1994), life sciences study strategies (Tanner, 2012), and computer simulation-based inquiry activities (Wang, *et al*., 2018). Goal orientation (Pintrich, 2000) and motivation (Pintrich, 2003; Pintrich, 2004; Pintrich & Zusho, 2007) are emphasized in students' self-regulated learning. Patterns of regulation were found different for higher performing students than for lower performing students (Stewart *et al.,* 2016).

When students regulate themselves in learning, their regulations are often implicit. There are artifacts, such as surveys, interviews, and reflective journals that can externalize self-regulation *after* it has happened (e.g., Dignath *et al.*, 2023). Still, planning, supervising, and evaluating can be automatic and develop without conscious reflection (Brown, 1987). When students are checking and regulating their own thinking and learning, it is often processed in their minds.

*Metacognition and Regulation in A Group Context*

In a group, metacognitive and regulative activities are extended beyond individuals and thus more likely to become explicit. For example, students can be more obligated to articulate their thinking to each other. They can be more aware of each other's thinking and more likely to respond to it. Goos *et al*. (2002) study the phenomenon of socially mediated metacognition in senior-level secondary math activities. They analyze transcripts of small group problem solving and examine collaborative zones of proximal development created through students' interaction with peers of comparable expertise. Investigating conditions under which such interactions lead to successful or unsuccessful problem-solving outcomes, they find unsuccessful outcomes are associated with students' poor metacognitive decisions and lack of engagement with each other's thinking. Successful outcomes are favored if students challenge and discard unhelpful ideas, actively endorsing useful strategies. Goos *et al*.'s work reconceptualizes metacognition as a social practice, laying a foundation for research in metacognition in social contexts. Van De Bogart and his colleagues apply Goos *et al.*'s framework of socially mediated metacognition and analyze audiovisual data from think-aloud activities in which students work in pairs to diagnose and repair a malfunctioning electric circuit (Van De Bogart *et al.*, 2017). The authors find that students engage in socially mediated metacognition in multiple key transitions during the troubleshooting process. Reciprocated metacognitive dialogue arises when students are collectively strategizing about which measurements to perform or reaching a shared understanding of the circuit's behavior.

When students work in groups, they have a shared task and are oriented by group goals, and motivation and emotion involve all group members. Regulations are operated at a group level (not only to oneself, but also to each other) and are therefore also likely to be more explicit. The areas of regulation are expanded to include multiple group members' cognition (e.g., Goos *et al.*, 2002; Van De Bogart *et al.*, 2017), behavior (Van De Bogart *et al.*, 2017), motivation and emotion (Järvenoja, 2010). Conceptualizing regulation in collaborative learning, Volet *et al.* (2013) claim that collaborative learning groups are composed of multiple self-regulating agents with distributed skills and knowledge, who may initially have incompatible goals. The group members jointly negotiate, coordinate, and regulate their collaborative pursuits to reach a shared understanding of the task, adopt effective strategies, co-construct knowledge, and work productively to complete the task.

The interplay between group members as self-regulating agents and their shared group tasks and goals provides an opportunity for researchers to examine the mechanism in which a regulatory operation in a group occurs. When a group member regulates the group, this regulation emerges from implicit thinking in an individual's mind into an explicit articulation and/or action to intervene, directing or adjusting the group activity. Conditions of transition from implicit to explicit can be challenges or concerns that group members face and feel they need to address collectively. Explicit discussions and changes in operation can be identified when an adjustment is proposed and pursued.

Regulation in a group setting encompassing self-regulation, co-regulation, and shared regulation (Järvenoja, 2010; Hadwin *et al.,* 2017). Self-regulation is students regulating themselves individually. Co-regulation is where self-regulation is supported by others. Socially shared regulation is when group members collaboratively regulate within the group. Hadwin *et al.* point out that regulation involves adaptively responding to new challenges, situations, or failure, thereby optimizing goal progress. To study regulative adaptation in groups, Hadwin *et al.* propose narrowing observations to periods in which challenges exist, and a regulatory response is warranted—an area that has not been well studied (Hadwin *et al.,* 2017).

*Socially Shared Regulations in Collaborative Science Learning*

As we pointed out in the introduction, science is a highly challenging discipline and offers intriguing contexts to study socially shared regulation. Research has examined various scientific aspects—such as problem solving (Goos, *et al.*, 2002), trouble-shooting (Van De Bogart *et al.*, 2017), experimental equipment (Bernhard, 2018), monitoring in a computer-supported learning environment (Borge *et al*., 2018; Sobocinsk *et al*., 2020, 2022), scientific argumentation (Lobczowski *et al*., 2020), learners' pre-conceptions (Lämsä *et al*, 2025), and students' conceptual uncertainty (Chen, *et al*., 2025)—when studying socially mediated metacognition and socially shared regulative learning.

Vauras *et al*. (2021) argues that in collaborative science learning, processes such as question posing, hypothesis generation and design, collection and synthesis of data, and the development and testing of models are inevitably becoming more complex than they are in an individually learning context. Therefore, complex reasoning and collective meaning making are continuous. Vauras *et al.* point out that there are research gaps in identifying and categorizing triggers and foci of socially shared regulation in the complex, collaborative science learning context, as well as in conducting more empirical studies.

We advance these lines of research to examine the instantaneous occurrences of students' regulatory moves in a collaborative science learning context. Grounded in our exploration of IMPRESS video data and connecting to existing theoretical frameworks, we propose our Emergent Explicit Regulation (EER) framework inviting researchers to attend, identify, and describe a regulatory move around its emerging moment. EER encourages researchers to examine closely the momentary externalization of regulation in a group and therefore gain a better understanding of socially shared regulative learning from a new, developmental perspective.

## Emergent Explicit Regulation (EER)

We introduce EER with an example that we identified from the IMPRESS video data early in our study where we established preliminary elements of EER (Cao *et al*., 2019; Ouimet *et al*., 2022).

**Brittany's example.** In an activity where four students (pseudonyms Pat, Justin, Jessica, and Brittany) are constructing a 3-D model to simulate the greenhouse effect. It is the second half of the class period, where students have spent the first half studying heat absorption of $CO_2$. Students are expected to build a final model based on what they have learned about heat reflection and heat absorption in the past three days. When the group works on building the model, Justin tries orienting a plastic fish tank in different ways for a while and it does not go anywhere. Pat then grabs the tank, fills in a thin layer of water, and brings it back to the table. Then Justin puts a sponge paddle in the tank claiming he is building a dam. Shortly after, Justin dumps the water and brings back the tank. There is no clear direction or discussion about where they are going. There are also a lot of small conversations while they casually build their model. As time passes by, the group still has not had a part of a model built. At this moment, Brittany, who has been chatting and observing with the group, says: "*Here. Better idea. Better idea*." She grabs the tank from the middle of the table to her side and adds paddles to the bottom of the tank while the others watch her. Jessica, sitting next to her, asks Brittany to explain her idea: "Say it before you..." Brittany refuses to explain and continues building: "No. I am going to do it. You guys are going to see." Justin then moves to the other side of Brittany and observes, which puts Brittany in a more centered position for the group. Brittany then builds the model in seven minutes and finishes it before the class ends.

In this example, the group is facing the *challenge* of spending too much time but not being able to build parts of a model. The regulation *emerges* when Brittany *explicitly* declares she has a better idea and grabs the tank to build a model herself. Brittany's *regulatory move* makes an adjustment of model building to potentially be more efficient, but at the cost of group collaboration and more opportunities for learning. This example shows the elements of an EER—a challenge, the context of emergence, explicit expressions, and a regulatory move, and the complex *effects* of a regulative move in a group learning context.

We continued finding more EER instances. Some are completed directly with one move. Others are completed gradually through multiple moves. The regulations also appeared to be in different psychological areas. To examine the EERs more systematically, we took into account what we have observed in the data formalized EER into an analytical framework.

*Formalizing EER*

In the EER framework, we define *regulation* as one or multiple regulatory moves a student can make to direct (Perkins & Salomon, 1989), or to adjust (Merriam Webster) the current process. To 'direct' is to make the group do things in a certain way when they have not. To 'adjust' is to make the group do things differently than what they are doing. Adjusting can be thought of as re-directing. Directing or adjusting can be made by one group member through one regulatory move, or by multiple group members through multiple moves. One student makes suggestions, poses questions about the direction to go or about the adjustment to make, and other students follow up and direct the activity or make the adjustment. In Brittany's example, Brittany makes a regulatory move to adjust the model-building activity. An example of multiple regulatory moves that we observed in our data is when one student makes the initial move proposing to use a longer measuring tool (longer than a meterstick) to measure the height of foams sprouting from a bottle of Diet Coke after they drop some Mentos tablets in the bottle, but the group does not have a longer measuring tool. A second student then follows up, brings two metersticks, tapes them front-to-end and uses it as a longer measuring tool.

The full EER framework is shown in Figure 1. Regulation is explained in this paragraph. Other elements will be explained in the following paragraphs.

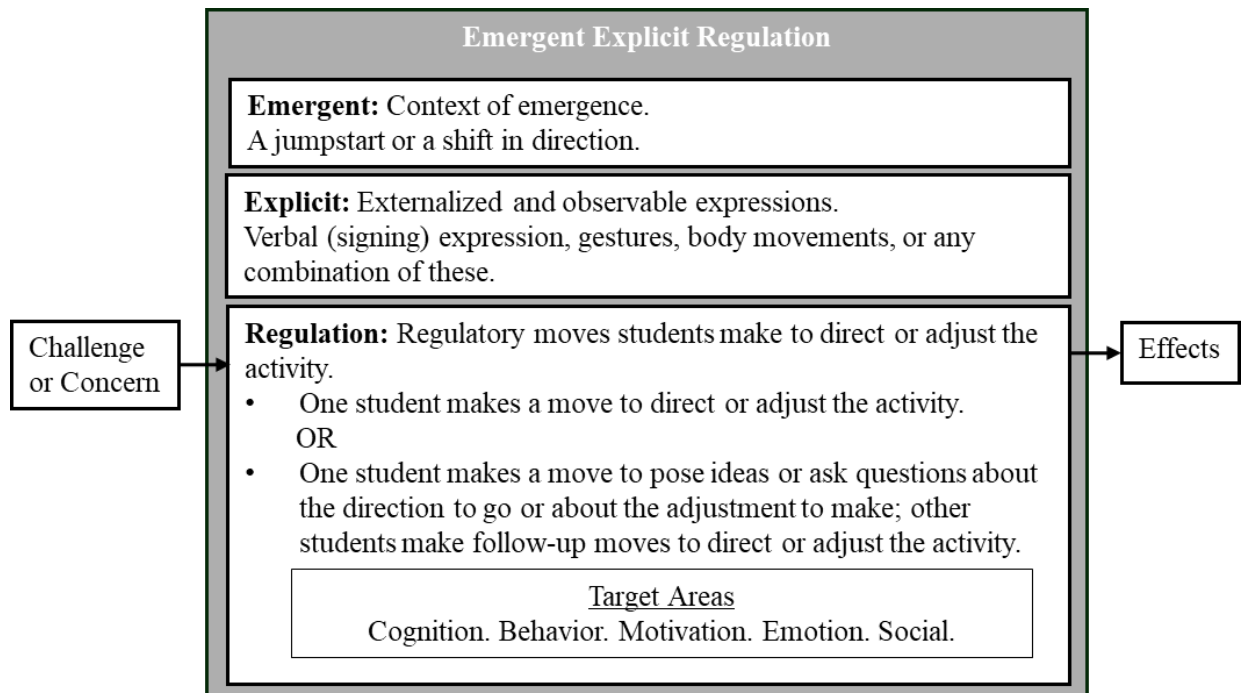

Figure 1. Emergent Explicit Regulation Framework

*Challenges or Concerns that Leads to Regulatory Moves*

The fact that one makes an explicit regulatory move indicates that there was a challenge or concern that triggered the move. The challenges or concerns are often implicit. We do not strictly distinguish a challenge from a concern. Whether it is a challenge that urges the group to address or a concern that only needs to be brought up to discuss depends on the group. In Brittany's example, the challenge was that the group spent a long time randomly trying and chatting and did not make any progress.

*Emergent: Context of Emergence*

The context of how a regulatory move emerges falls into one of the two kinds: a *jumpstart* (when the group is not actively talking or doing something, and someone breaks the silence to speak or act), or a *shift in direction* (when the group is in the process of talking about a topic or operating a task, and someone brings up a different idea, suggests or directs a different operation). The two contexts are related. One could argue that a jumpstart is also a kind of shift in direction (from silence), which is fair. We categorize the two contexts to more aptly notice and separate the context of the EER emergence—one is the group being silent/stagnant, and the other is the group doing and talking actively.

The context of emergence in Brittany's example is a shift in direction, because Brittaney's move emerges in the middle of other group members actively talking and putting materials in the plastic tank.

*Explicit: Externalized, Observable Expressions*

The regulatory moves are explicitly expressed in one or more ways: verbal expressions (some are signs in the IMPRESS context), gestures (nodding, waving hands, pointing to an object), body movements (grabbing or moving equipment, walking, standing up, sitting down, changing seats or standing position), or any combination of these. These expressions are observable and help us identify and describe an EER to better understand the content of an EER. Brittany's regulatory move is explicitly expressed verbally ("Here. Here. Better idea. Better idea.") and body moments (grabs the tank and builds a model).

*Regulation Target Areas*

As we have noticed, there are multiple areas where regulations can emerge and develop. To categorize these areas, we adopted the categories from Sobocinski *et al.*'s framework (2020) that the authors categorize students' areas of monitoring. Those areas include cognition (thinking, often involving task understanding), behavior (operation, often involving task enactment), motivation (attention and engagement), and emotion (affects). We added a new category, social (interaction and tensions between group members) based on our observations in our data. One EER instance can address one or multiple target areas. In Brittany's example, the regulation target areas are cognition (idea of how to build a model) and behavior (building the model).

All five EER target areas with descriptions and examples are shown in Table 1. The examples are based on real cases in our data but modified for brevity.

Table 1. EER target areas and demonstration in scientific inquiry.

| Target Area | Description of target area | Example |
|---|---|---|
| Cognition | Regulating thinking and ideas<br>• Understanding task<br>• Designing and ideas<br>• Recalling content knowledge | The group is designing a physical model of the greenhouse effect and will measure the temperature change in the model when it is heated. They are attaching a thermometer to the rim of the model. A student jumps in and says: "We should put the thermometer inside since we need to know the temperature inside." (i.e., designing and ideas) |
| Behavior | Regulating operation<br>• Enacting task<br>• Gathering materials<br>• Operating equipment | The group needs to heat up the model for 10 minutes. One student is monitoring the time but does not sit next to the heating lamp switch. When 10 minutes is up, this student tells the other student who sits next to the heating lamp switch: "Turn off the light." |
| Motivation | Regulating engagement.<br>• Redirecting off-task chat<br>• Leveraging low interest | The group are distracted and check their phones, not getting work done. One student says to the group "Let's talk about what we need to do here." |
| Emotion | Regulating emotional state.<br>• Comforting<br>• Cheering up | A student gets upset because the data is not shown as expected. Another student comforts the first student: "It's okay." |
| Social | Regulating interaction between group members is important.<br>• Managing conflicts<br>• Building rapport | Two students are arguing intensely. Another student tells them to pause. |

*EER Effects*

The effects of a regulatory move or moves are manifested in what the group does next due to the adjustment they make. It is possible that there are no effects even when regulatory moves had been made—the group did not take up the expressed regulation and continued doing what they had been doing before. The latter could happen for various reasons (e.g., someone being stubborn, the suggested adjustment being too complicated to implement). In Brittany's example, the effects are that Brittany built the model by herself in seven minutes, but other group members were not involved in the model building process.

## EER in Collaborative Inquiry

The EER framework was born in our study of students' metacognition in learning science. We are interested in situating EER in the context of collaborative scientific inquiry learning and focus our empirical study on unpacking students' spontaneous regulations in the process of inquiry science.

According to Vauras *et al.,* (2021), three core aspects characterize the nature and contexts of science learning. *Authenticity* (embedding the learning in learners' everyday world and the practice of the discipline), *collaboration* (learners sharing and contrasting of ideas with other individuals sharing similar aims), and *inquiry* (students engage in problem-solving activities that require planning, synthesis, and evaluation skills as well as relevant domain specific content knowledge). We adapt these three cores and elaborate on them to have a comprehensive set of features describing collaborative inquiry learning in Figure 2.

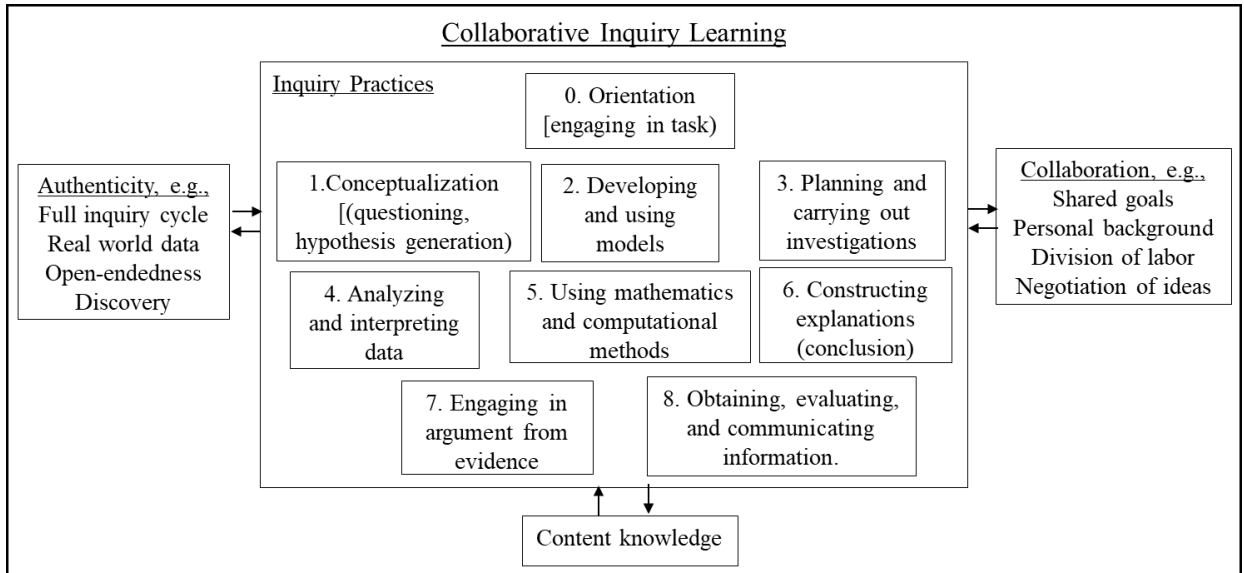

Figure 2. Collaborative Inquiry Learning

In Figure 2, the central box contains the elements of *inquiry practices*. In Vauras *et al.*'s description of *inquiry* mentioned in the previous paragraph, inquiry includes both practice and content knowledge. We keep inquiry practice in one box and separate content knowledge outside of practice. This separation aligns with the Next Generation Science Standards (NGSS) which separate the strand of *practice* ("science and engineering practice") from the strand of content ("disciplinary core ideas") (NRC, 2013).

We primarily adopt the NGSS's eight science practices (NGSS Appendix F) to elaborate the inquiry practice (numbered 1-8). Additionally, we are informed by Pedaste *et al. al*'s systematic literature review (2015) summarizing inquiry-based learning into five phases: orientation, conceptualization, investigation, conclusion, and discussion. Their last four phases overlap with the eight practices in NGSS. Their first phase—orientation is not explicitly reflected in the NGSS science practices. Orientation, according to the authors, involves introduction, anchor, finding one's topic and learning challenges, and engagement. We recognize this phase as additional practice and add it as practice zero. We also use "conceptualization" for practice 1, instead of "asking questions" to include both asking questions and generating hypotheses. A description of all 0-8 inquiry practices is provided in Appendix A.

Authenticity and collaboration are the other two core aspects in science learning, according to Vauras *et al*. (2021). We organize them on the sides of inquiry practices. For authenticity, we derive from Vauras' definition and include open-endedness, real-world data, and new discovery as examples. For collaboration, we also start with Vauras' description and include personal background, division of labor, shared goals, and negotiation of ideas as examples. Authenticity, collaboration, and content knowledge are permeant though all inquiry practices.

Having this comprehensive characterization of collaborative inquiry learning, abbreviated in C-Inquiry, we can situate EER in this context and perform more systematic analysis. For example, in Brittany's example, the EER addresses inquiry practice 2: developing a model.

Brittany's EER of all components is illustrated in Figure 3. For each EER instance, we can construct an EER diagram with its specific components.

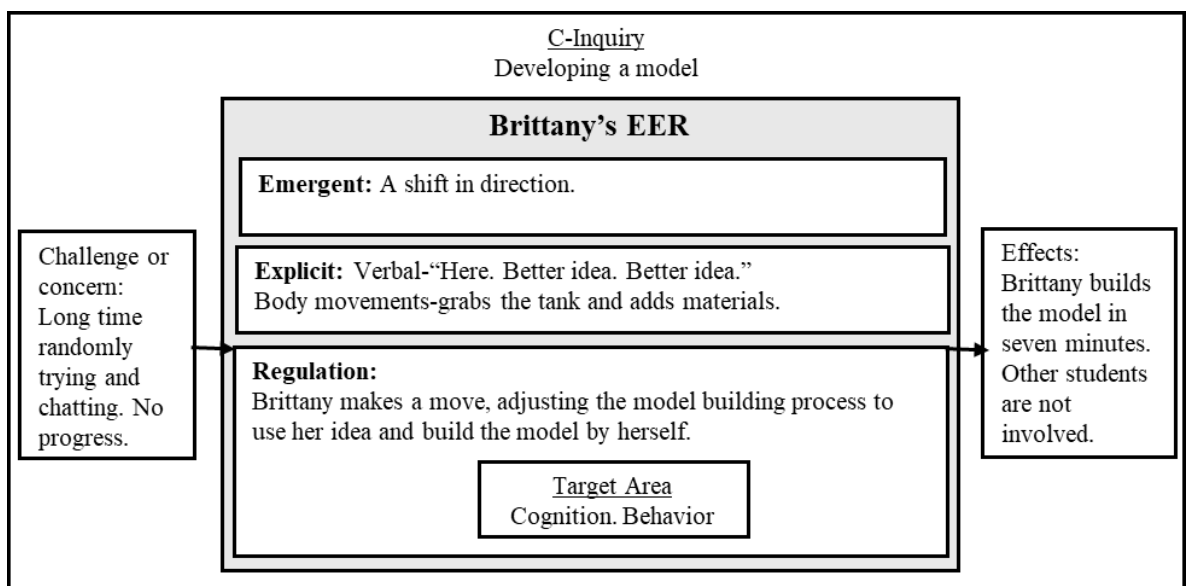

Figure 3. Brittany's EER.

In the following, we present a case study (Creswell & Creswell, 2017) in which EER is applied to analyze a group of four students using mixed communication methods completing a sequence of two hands-on, inquiry-based activities in two consecutive days. Through this case study, we further demonstrate EER's role in unpacking the in-the-moment regulatory moves made by group members toward achieving shared learning goals in collaborative scientific inquiry learning.

## Methodology

*Context*

The IMPRESS classroom layout is shown in Figure 4 (top view). Students (purple circles) were in groups of four sitting at one of the five tables (green shapes), where a camera (blue icon) is installed to record video data. The circled table at the bottom is where our case study group sat during the activities we analyzed.

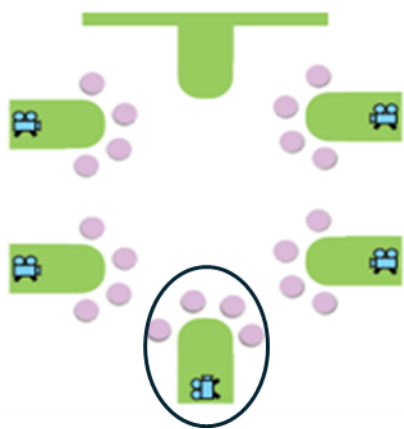

Figure 4. IMPRESS classroom layout. The Circled group is the case study group.

The IMPRESS program has a sequence of hands-on experiments modeling the greenhouse effect. The sequence is designed to develop students' understanding of climate change and to experience the iterative process of constructing an initial model, testing thermal effects of specific materials by collecting data while heating and cooling the model with different materials, and iteratively revising the model based on what they learned from the temperature data. This sequence spreads from Day 2 through Day 5 of the ten-school-day program. The activities are described in Table 2. Brittaney's example described earlier was on Day 5, part (2) Final model construction.

Table 2. Greenhouse Effect Model Construction Sequence.

| Day | Activity |
|---|---|
| 2 | **Initial model construction.** Students are provided materials (a plastic fish tank, a black mat, a white foam, a blue felt, tin foil, baking paper, plastic wrap) to construct a model representing the Earth atmosphere and the greenhouse effect. Students collect temperature data while heating up the model by a heating lamp for 25 minutes and cool down for 25 minutes after turning off the lamp. Temperature data are collected though thermometer probes connecting to a shared laptop, with software installed to show the temperature on the screen. |
| 3 | **(1) Thermal absorption.** A fish tank is set upside-down with a black mat at the bottom. Students measure the temperature in the fish tank as it heats up and cools down for 25 minutes, respectively. One round is with a black mat. The other round is without a black mat.<br>**(2) Model revision.** Students take what they learned from the black mat experiment and rebuild their model. They then heat up, cool down their model, and collect temperature data. |
| 4 | **(1) Thermal reflection (Albedo Effect).** A fish tank is set upside-down with white foam at the bottom on top of a black mat. Students measure the temperature in the fish tank as it heats up and cools down for 25 minutes, respectively. One round is white foam. The other round is without white foam.<br>**(2) Model revision.** Students take what they learned from the white foam experiment and rebuild their model. They then heat up, cool down their model, and collect temperature data. |
| 5 | **(1) Thermal absorption—$CO_2$.** Two flasks of water and a small bag of cold relief tablets are given to each table. Students add four cold relief tablets to one flask. They heat up the two flasks at the same time for 25 minutes and cool down for 25 minutes.<br>**(2) Final model construction.** Students take all they have learned and build their final model. They heat up, cool down their model, and measure the temperature. |

*Data Source and Case Selection*

Our case selection is guided by our research questions and the available data. The IMPRESS program ran for four summers from 2014 to 2017. The classroom videos are our primary data. Additional data includes the program schedule, students' demographic data, classroom setup figures, lesson plans, video catalog, field notes by research assistants, and scanned copies of students' reflective journals. Within the four years, the 2015 and 2016 program data are best documented. The 2016 IMPRESS students all used speech to communicate, missing one important communication feature in the IMPRESS student population. We thus narrowed down to 2015 data. The 2015 IMPRESS students originally formed small groups on Day 1 of the program and stayed in their small groups through Day 3. On Day 4, students regrouped and stayed in the same groups for the rest of the program.

In the 2015 data set, we looked for a typical case (Seawright & Gerring, 2008), a group that (1) engaged in inquiry-based, hands-on science activities, (2) worked collaboratively on the group task, (3) had rich interactions where EERs can be more accurately identified when they occur, (4) used mixed communication methods, and (5) the videos and audio are clear enough to perform qualitative analysis. Through careful examination, we found the following group in the Day 4 and Day 5 greenhouse construction activities meeting our criteria. There are 150 minutes' videos recording this table on their Day 4 greenhouse model construction activity and 165 minutes' videos recording this table on their Day 5 greenhouse model construction activity.

The students' demographic information is in Table 3.

| Pseudonym | Gender | Ethnicity | DHH Communication Method |
|---|---|---|---|
| Ashley | F | White | Signing only |
| Grace | F | Asian | |
| Jill | F | White | Speech |
| Sara | F | White | |

The equipment set up and students' searing during Day 4 and Day 5 greenhouse model construction activities are shown in Figure 5.

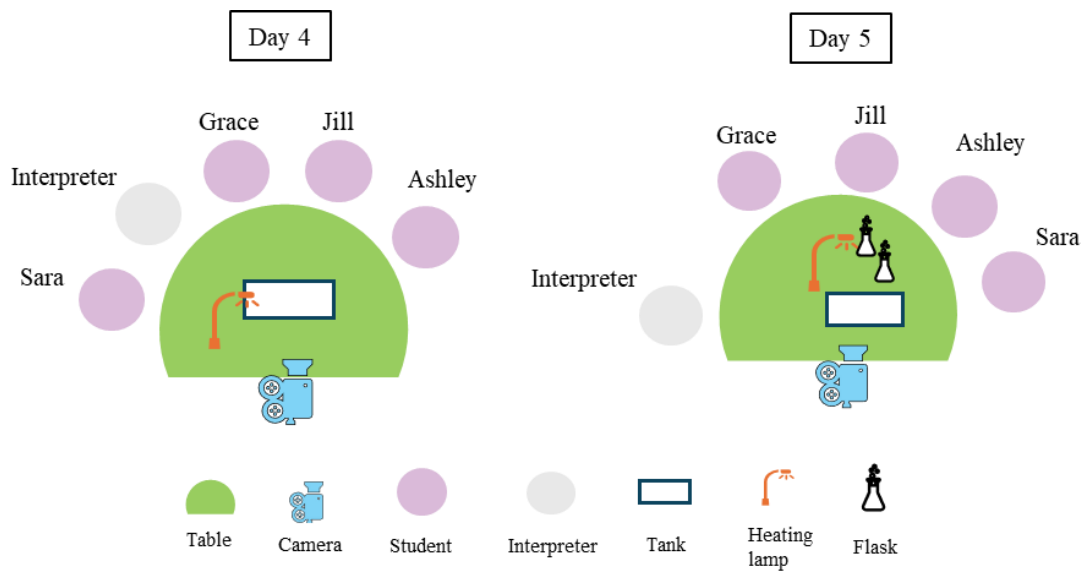

Figure 5. Case study group table set up and seating on Day 4 (left) and Day 5 (right)

*Data Analysis*

Three of the authors of this paper analyzed the videos fully. The first author trained the other two authors on the EER framework and methods of analyzing video data qualitatively. Then, the three authors iterated the process of (1) independently watching the videos, (2) independently identifying EERs and describing their elements; (3) meeting, discussing, and resolving any disagreement on their individually identified EER instances; and (4) repeating (1)-(3) until the identified EER were all agreed. In (1) and (2), we transcribed the video snippet around each identified EER instance, including timing, students' verbal expression, DHH student's signing and interpreter's verbal translation, students' gestures and body movement, and relevant classroom environment. We also paid attention to the short period before and after an EER instance that we considered helpful in understanding the EER and jotted down relevant observations. Using this iterative method, the three authors established a pool of all-agreed EER instances.

All five authors then from the pool of EER instances selected typical EER instances to present in the results of this paper. The other two authors were familiar with the videos through previous work in this project. When selecting typical EER instances, we drew the ones where all five authors agreed on all the elements in an EER defined in Figure 1, based on reading the transcripts and, when needed, rewatching the relevant part of the video. We considered selecting cases from different target areas, from the inquiry practices designed in the activities, more directly contributing to completing the group task.

In addition to the videos, we also checked the program documents and the four students' journals to help us understand the broader context and triangulate our interpretation of the video data.

*Validity and Reliability*

To ensure validity, we base our EER framework on the literature of socially shared regulation (e.g., Hadwin *et al.,* 2017). We adapt target areas from established scholarly work (Hadwin *et al.*, 2017; Sobocinski *et al.*, 2020 & Sobocinski *et al.*, 2022) and reference multiple sources to determine elements in collaborative inquiry (NGSS, 2013; Pedaste *et. al. al*, 2015, Vauras *et al.*, 2021).

For reliability, we rely on observable phenomena (speaking, gesture, and body movement) to identify the EER instances. All authors have worked together to develop and refine the EER framework through preliminary work and multiple iterations. As described in data analysis, multiple authors analyzed the videos for multiple rounds in different degrees of detail to determine the typical EERs to be presented in the results.

## Results

During the two days, the four students formed a new group, completed the instructed experimental tests (Albedo effect and Thermal absorption $CO_2$), and built two versions of the greenhouse effect model. Additionally, they tested materials that were not required, but that they were interested in incorporating in their model. Moreover, the group observed unexpected data trends, adjusted data recording details, and came up with model representations based on their data. In this section, we will present typical cases of EER in each target area and elaborate how these EER instances contributed to completing the group task and achieving group goals in collaborative inquiry learning. We will show how the students'

- Social EERs contributed to forming a new collaborative group;
- Motivation EERs contributed to re-engaging the group in carrying out the experiment when they are distracted;
- Behavior EERs contributed to organizing a flexible and efficient division of labor in carrying out the experiment;
- Cognition EERs contributed to adjusting the interval and range to take numerical data and integrating group members' diverse ideas to construct a satisfactory model.

The EERs elaborated in this section all emerged when the students were collecting data. However, the challenge or concern in each instance varied, so did the direction and adjustment the group was moved into by the regulatory moves. Social EERs directed the group out of inquiry practice to a social conversation. Motivation EERs re-directed the group back on task to focus on the current step when some group members were distracted. Behavior EERs improved the ways of managing data to make the experiment better operated. Cognition EERs appeared to shift the group's activity from passively waiting for data to actively adjusting data collection method or planning for next steps such as collecting ideas of building a model.

In the following, we will present each typical EER instance with the transcript of the video clip when the EER occurred and the immediate context. We also describe the broader classroom context and the activity information, as well as some excerpts from students' journals when they are helpful to illustrate the EER. In the end of each instance, we summarize the EER in accordance with the EER framework shown in Figure 1, specifying each element. We present the cases in chronological order to make the progress of the two-day activities easier to follow.

*Social EERs Contribute to Forming a New Collaborative Group*

We found social EERs on Day 4 when the group was newly formed. At the beginning of Day 4, after the students came in, they were instructed to walk around and work with someone they had not worked with before. Grace, Jill, Sara, and Ashley sat at the same table and formed a group. Grace and Jill were in the same group from Day 1-Day 3. Sara was in a different group. Ashley was in a different group.

The instructor then announced the task for the day. Each group was given a model (a plastic tank upside down covering a piece of white foam on a bigger black rubber mat on the table. The goal of this part is to test the effect (the temperature change) of a reflective surface (the white foam) on the simulated Earth atmosphere (inside the fish tank) when the model heats up by a heating lamp from the top and then cools down after the heating lamp is turned off. After that, each group would rebuild their model based on what they have learned thus far.

The group turned on the heating lamp. Sara set the timer. Temperature data were automatically collected via a computer program through a thermostat probe placed in the tank. The group was sitting around the table and quietly waiting. They occasionally checked their laptop but most of the time they stared in silence. Then, Ashley broke the silence and signed Sara, the interpreter translated.

- Ashley signs and the interpreter says: (to Sara) What is your name?
- Sara: "Um, my name is…" [orally says her name and signs the spelling of her name.]
- Ashley signs to repeat the letter Sara signs.
- They then sign together, and Sara speaks repeating the letters.
- Ashley nods when they finish signing.
- Ashley then turns to Grace, who is a speaking student but appears to know how to sign the letters. Grace tells her name orally and signs the spelling of her name to Ashley, and Ashley repeats the signs.
- Ashley then turns to Jill, who is a DHH student but speaks to communicate. She also appears to know how to sign the letters and signs the spelling of her to Ashley. Ashley repeats the signs.
- Finally, Ashley signs her own name, and the interpreter translates it to the group.

This is a social EER where the challenge was that the group sat at a table and waited for a long time for data collection when the members did not yet know each other. Ashley's regulatory move was a jumpstart, breaking the silence and asking everyone's name by signing—initially interpreted by the interpreter, then directly signed between the group members, and lastly conveyed again by the interpreter. The effect was that the students got to know each other and found something to do while waiting. This EER directed the group temporarily out of the inquiry practice to chat with each other. This EER is illustrated in Figure 6.

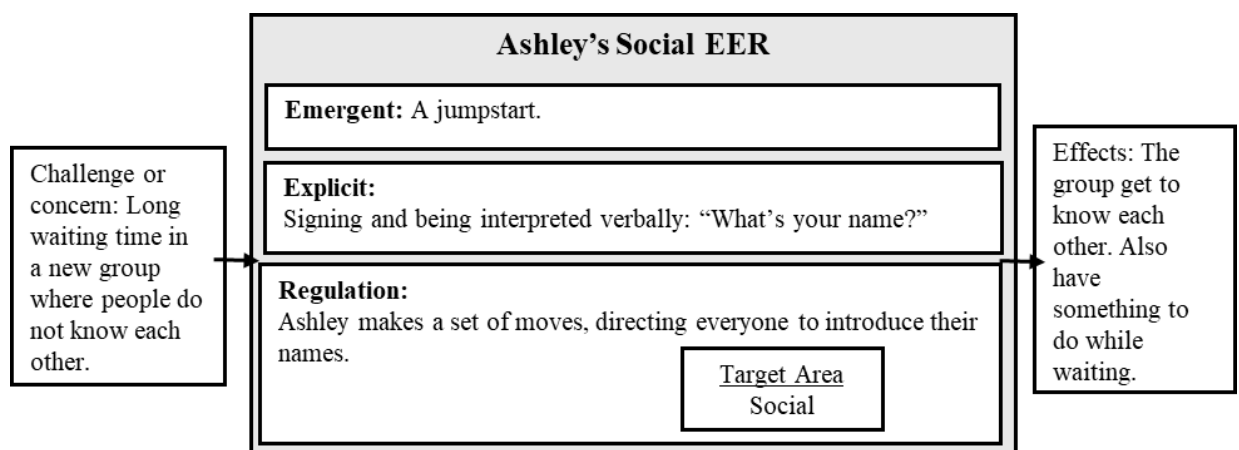

Figure 6. Ashley's social EER

Later in the activity, when the group waited to heat up and cool down different versions of the model, there were similar social EERs. For example, Grace asked the group what they did over the weekend (Day 4 was a Monday). They chatted about that. Ashley signed to ask everyone about their majors. The group went around, and each student introduced their majors. Also, Sara asked Ashley to teach them how to sign. Ashley taught the group to sign some words.

These social EERs kept the new group engaged in their group, even though the social conversations were not directly about the experiment. These social EERs directed the four students to form and maintain a group for collaboration. Ashley, Grace, and Jill all mentioned their positive feelings of being in a new group in their journals.

> I changed table and ended up with an entirely different group which is fine with me since I can get along with the hearing people just fine. [Ashley, Day 4]

> The action of "switching the tables, and partners" brought perspectives and point of view of how our group approached the task but also allowed us to be able to work and interact with each other in a different way. [Grace, Day 4]

> Having to move to a new table definitely made things more interesting. There was a different dynamic and different ideas. [Jill, Day 4]

*Motivation and Behavior EERs Contribute to Adjusting the Equipment, Re-engaging Distracted Group Members*

On Day 4, Sara and Ashley made regulatory moves together to address challenges in both motivation and behavior areas. During the experiment, a thermometer was set up on Sara' side of the table, facing Sara, and took temperature data from inside of the tank- the degrees being shown on the computer screen. Another group sitting next to the door made a loud noise talking to the instructor. Grace and Jill were distracted by their conversation and turned to look the other way. Sara was also briefly distracted but soon turned back to the table. Then the interpreter signed to Ashley about what the other table was talking about, Grace and Jill looked at the interpreter. At this moment, Sara made a move and picked up the thermometer, looked at its screen, and asked if there was a way to set the thermometer unit in Fahrenheit (which indicates that the unit was originally set in Celsius).

-Sara: (picks up the thermometer) Is there a way to set this to Fahrenheit?
-Grace and Jill turned back to face the equipment.
-Nobody responds to the question. There is a pause. The group left the thermometer as is.
-Ashley has a sip of water. She puts down her water bottle and raises her right hand with her index finger pointing up (a hand gesture like when the interpreter interprets "Hold on.")
-Ashley reaches her hand to Sara's side and grabs the thermometer, turns it upside down, finds the switch, and switches it to the other side. She looks at the monitor of the thermometer and returns the thermometer to Sara's side on the table.
-Sara looks at the thermometer and nods.

This is a motivation and behavior EER in carrying out investigation, completed collectively by Sara and Ashley. The first part was regulating motivation (group members' attention and engagement), and the later part was regulating behavior (adjusting equipment). In this example, the challenges or concerns were that the group members were distracted, and the thermometer was not set in the expected unit. Sara made a regulatory move while the group was not actively doing the experiment and verbally asked the group to the group. This made Grace and Jill turn back, and Ashley appeared to also listen to Sara then. The group initially did not make an adjustment to the equipment. After a pause, Ashley followed up and made a move to switch units. The effects of Sara and Ashley's EER are that the group was re-engaged in the experiment, and the thermometer's temperature read in Fahrenheit. This EER is illustrated in Figure 7.

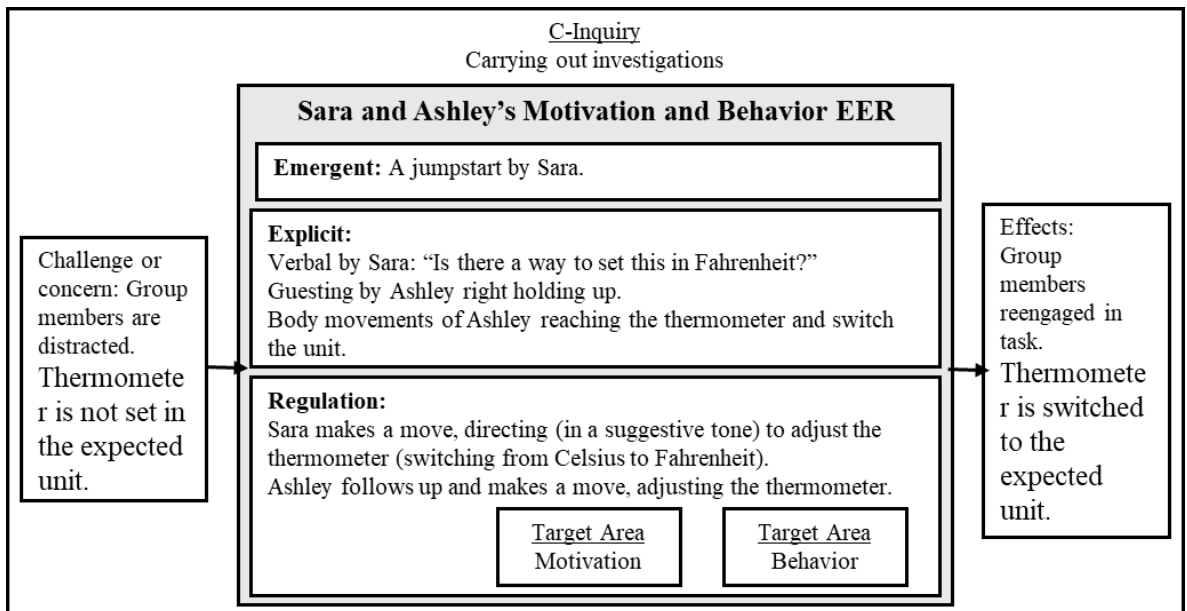

Figure 7. Sara and Ashley's motivation and behavior EER

Jill mentioned in her journal on Day 4 that the classroom has a lot of distractions.

> I feel like there were a lot of things that distracted me, like my phone or a conversation at the next table. Blocking out those distractions and removing the sources of distraction that I can would help. [Jill, Day 4]

Despite the distractions, this group overall managed well to stay engaged in the experiment. In the two days, we only found a couple of instances where a group member regulated motivation. There is another case on Day 5. Sara and Ashley sat side by side facing the thermometers. Grace could not see the thermometers but needed the temperatures to be read to her. Grace noticed that Sara was distracted and turned to the table behind her, so she asked the interpreter "Can we have the temperatures now?" The interpreter signed this to Ashley. Ashley patted Sara's shoulder and made her turn back and read the temperatures to Grace. In this EER example, the regulation was also in both motivation (back from distraction) and behavior (reading the temperatures).

*Behavior EERs Contribute to Organizing a Flexible and Efficient Division of Labor*

There are also behavior EERs that directed the group members to work collectively, form an efficient division of labor, and stay organized in carrying out the experiment. At the beginning of Day 5, the instructor announced the tasks for the day. First, testing $CO_2$ absorption with the given equipment—each table was set up with two flasks with water and a bag of cold relief tablets (which will release $CO_2$ when dropped in water). Students were instructed to drop four cold relief tablets in one flask, heat up and cool down the two flasks simultaneously, and record the temperatures. Next, they were expected to take all they had learned in the past few days and so far, that day to build their final model of the greenhouse effect.

When the group got started, Sara handed the cold relief tablet bag to Ashley asking if she would drop the tablets in one of the flasks, which Ashley did. Grace then turned on the heating lamp. At this moment, Jill suggested the group write things down, which led the group to collaboratively carry out the following experiment and also record the data manually in their notebooks rather than solely relying on the computer program.

> -Jill: "carbon dioxide is a reactive gas, so as soon as you put those in there and put the top back on."
> -Sara (to Ashley): "do you want to add this?"
> -Ashley takes the package and gestures four fingers to Sara. Sara nods. Ashley puts four tablets in one flask and puts the cap back on.
> -Jill: "We should write stuff down."
> -Sara takes her backpack and notebook and moves to the other side of the table where she can read the thermometer directly (see Figure 8).
> -Ashley points to Sara the two thermometers in the flasks and signs, translated by the interpreter: "Here. It's all in Fahrenheit."
> -Sara reads the thermometers: "So this one is…" [reads the temperature and writes on her notebook]

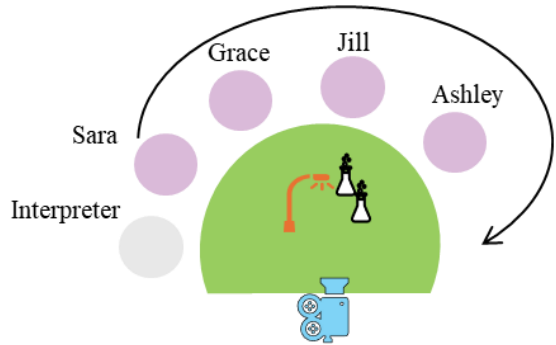

Figure 8. Sara changes seat.

Following some more discussions about who does what and how to do it during when they recorded the first couple rounds of temperatures, the group formed a division of labor that lasted for the rest of the activity: Sara read temperatures. Jill and Grace both kept track of time and told Sara when to read the temperatures. Grace, who then became the one who sat closest to the lamp, turned the lamp on and off when needed. Sara, Grace, and Jill all recorded data in their notebooks. Ashley watched the temperatures on the laptop screen but did not write notes in her notebook.

This behavior EER, started by Jill and completed by the group, had a lasting impact on how the group carried out the experiment. The challenge or concern was that, according to Jill, the group did not carry out the experiment well organized (see her journal quotes later). Jill made a shift in direction in verbal expression, followed by Sara and Ashley in their body movements, signing, and gestures. The group, since then, had a stable division of labor and carried out the investigation more organized. Writing data down in their notebook also had an impact on the way they viewed the data, recognized and compared data trends, and used their observations to build and justify their final model (see more details later). This behavior EER is illustrated in Figure 9.

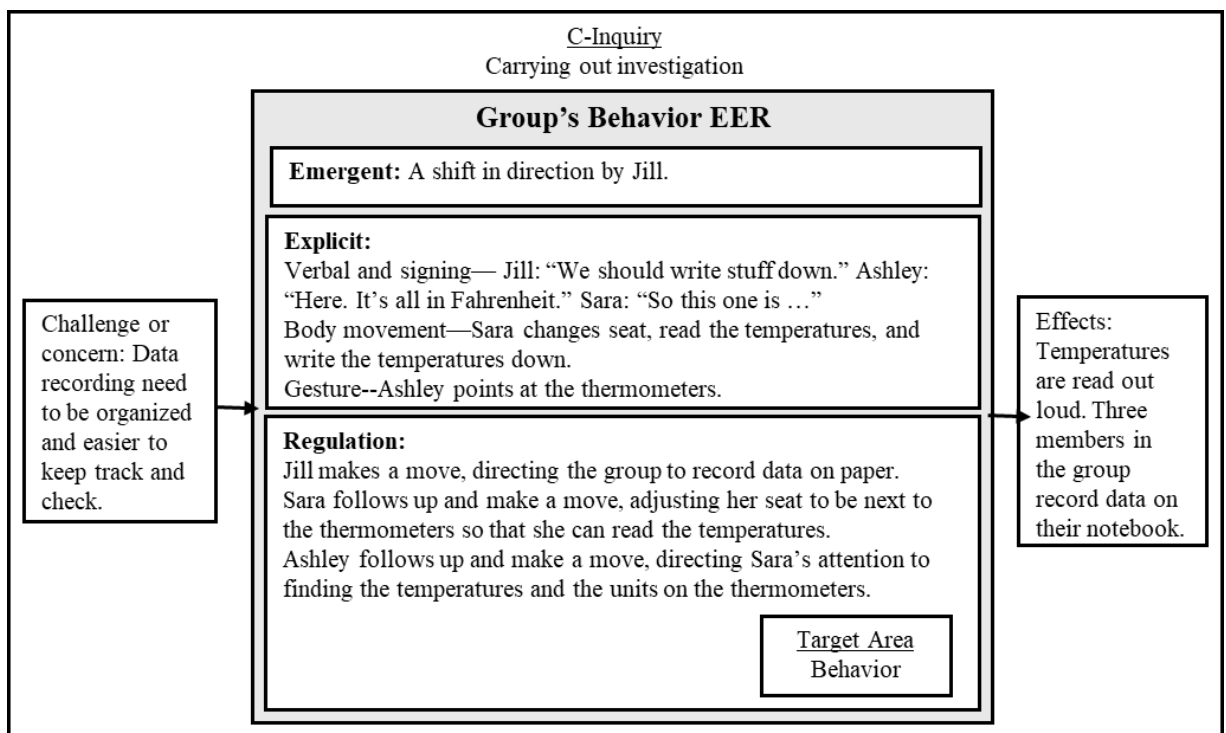

Figure 9 Group's behavior EERs.

The group's ways of carrying out the experiment appeared to have to do with the physical set up of the equipment—who sat next to which piece of equipment (heating lamp, timer, thermometer, and flasks). Staying organized and carrying out the experiment more efficiently was what the behavior EERs directed the group into. Altering the physical positions (of people or of equipment) was one demonstration of the behavior regulations. In the previous example, Sara picked up the thermometer from the table. Ashley grabbed the thermometer from the other side of the table and returned it after switching the unit over-rode. In this example, when Jill said, "We should write stuff down," Sara moved her seat and changed her role to be reading the temperature (she had been in charge of turning the lamp on and off on Day 4). Grace did not move, but because Sara moved away from the heating lamp, Grace became the closest to the lamp and was in charge of turning the lamp on and off on Day 5.

This EER directed the group to carry out their experiment and record their data more organized. Jill's journal reflected this change between Day 4 and Day 5.

> Another thing I noticed was that our group today really wasn't very organized. We would jump from one test to another without even really completing an experiment. [Jill, Day 4]

> Our group organized our data, so we were able to determine what materials represented what. [Jill, Day 5]

Sara, who changed seats on Day 5 and became in charge of reading temperatures and recording data, also wrote in her Day 5 journal, in her itemized (in numbers) reflections, about her positive feeling of taking control of recording data.

> #3 I really think taking control and analyzing the information for myself instead of someone telling me their interpretation of how it works is better for me.
> #4 Taking and recording the data for myself. [Sara, Day 5]

*Cognition EERs Contributing to Adjusting the Interval and Range of Recording Data*

The group watched their data more closely on Day 5, partially due to Jill's behavior EER to make them "write stuff down," and partially because the group noticed unexpected data trends at the beginning of Day 5. According to the data recorded in Sara's journal, right after the lamp was turned on, the plain water was 98.7 °F, and the water with the four tablets was at a lower degree—89.4 °F. During the first five minutes of heating, the plain water soon reached 132.9 °F, while the water with the tablets only reached 107.2 °F (see Figure 10, left, top section).

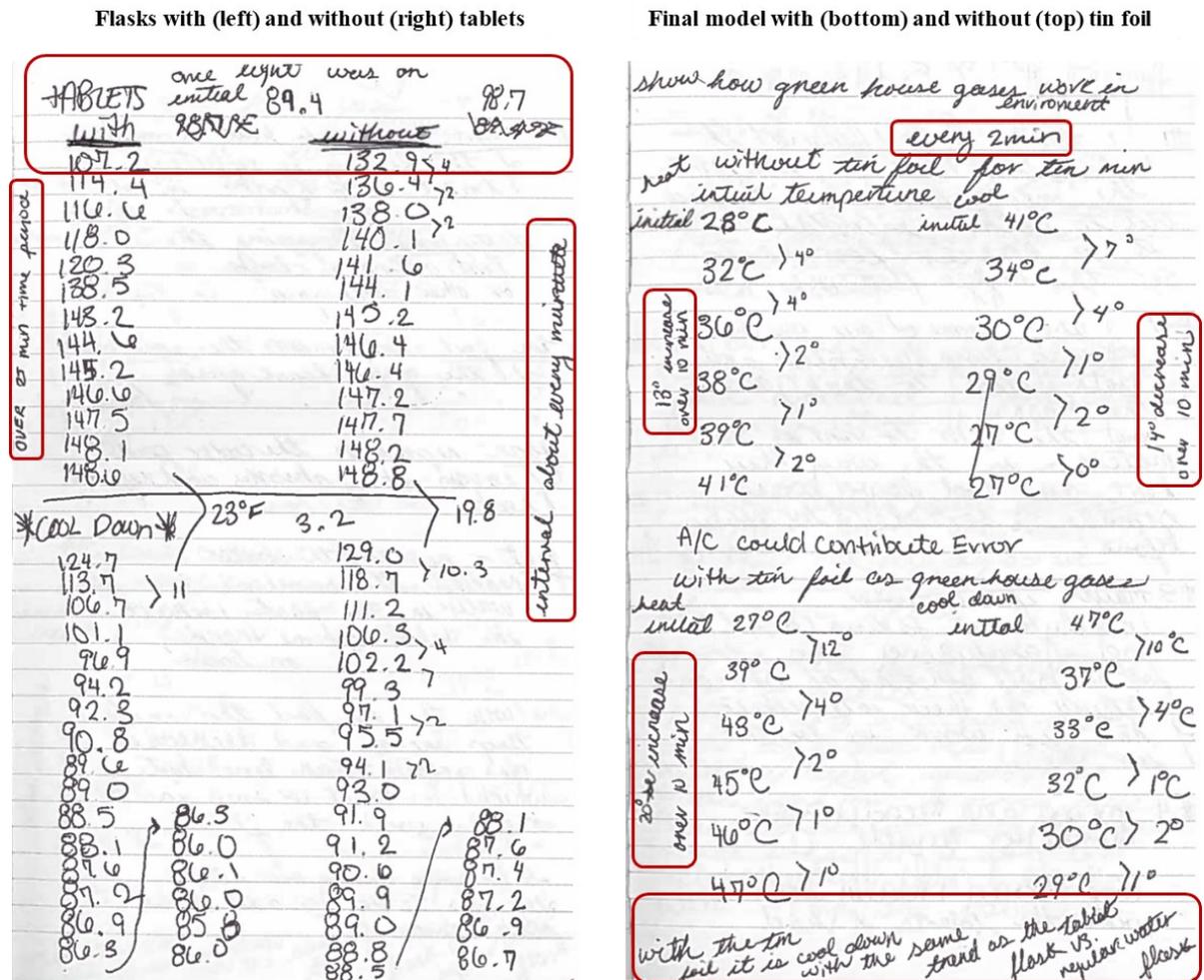

Figure 10. Data recorded in Sara's journal. Left: Temperature data on the two flasks of water with and without the tablets. Right: Temperature data on the final model with and without the tin foil. Circled parts of Sara's notes are mentioned in the paragraphs.

The phenomenon might have contradicted the group's knowledge of $CO_2$ being the Greenhouse gas which should have made the water warmer than plain water. One hint of this is that in Sara's journal she had recorded the two temperatures flipped but then crossed them off (see Figure 10 left, top section). Later in the video, the group had verbally checked and confirmed several times that they recorded the temperatures correctly. To make sure the group did not simply make a recording mistake, we checked the program lesson plan of this activity. Data in the lesson plan from the experiment the TA did prior to the lesson showed that the flask with tablets had a lower temperature overall compared to the flask of plain water.

Jill could be the first one who noticed the plain water being warmer, according to the following transcript, in the first five minutes of heating.

-Ashley describes the temperature trends: "This one's skyrocketing and the other one's just taking its own sweet time."
-Jill: "I wonder what is between the water and the bottle?" (Here, Jill might have noticed that the plain water being warmer, so she asked this question.)
-Ashley: "What? What do you mean between the water and the bottle?
-Jill: Well, I mean... (inaudible)
-Sara (to Jill): "Wait, so tell me when it's been, like, four or five minutes so I can write down the time."
-Jill: "Should we start recording it more often, like, maybe every two minutes?"
-Sara: "We can do it every minute, I guess."

The group record data everyone minute since then (see Figure 10 left, right side note).

We categorize this EER as a cognition EER, although we recognized that the regulation also made an impact on the group's behavior (recording data). The reason that triggered Jill to suggest adjusting the time interval of recording data was the radical change of the temperature within the first five minutes and the unexpected trends in the two flasks. The five-minute interval was not fine enough to capture the details of the temperature change. Monitoring numerical data at a higher frequency and learning more details about the rate of change in relation to a physical phenomenon involved observation, mathematical thinking, and data interpretation, and more. Therefore, this regulation demonstrated a cognitive challenge, and the adjustment was towards a better understanding of the data and the phenomenon. This EER is shown in Figure 11.

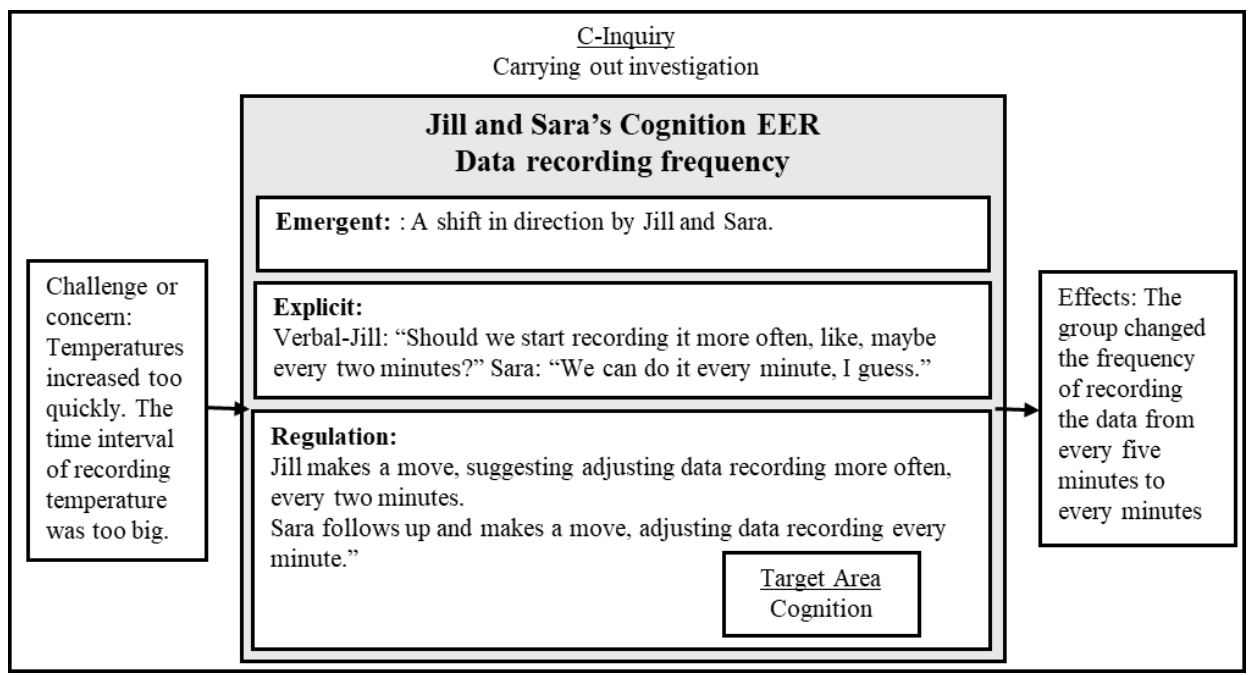

Figure 11. Jill and Sara's cognition EER in data recording

While the flasks of water were heating up for the next 20 minutes after they changed the time interval of recording data, the group were verbally keeping track of time, reading, recording, and comparing data. Ashley appeared to be intrigued and enthusiastically asked several questions intermittently about why the data trends were like that. Jill proposed some guesses, saying maybe there is something in there (in the plain water), but the group did not follow up with that. They were, however, alerted by the data trends and became more observant of the data and more aptly comparing data. They subsequently made some more adjustments regarding the time range for recording data. After they heated up and cooled down the flasks for 25 minutes each, as instructed, they noticed that the temperatures had leveled off after around 10 minutes. When they heated up and cooled down their final model, they changed the time range to 10 minutes each (see Figure 10 left, left side note and Figure 10 right, both sides' notes). It also seemed that they had changed the interval to be every two minutes when recording temperature for their final model (see Figure 10 right, upper part), but we did not hear that part of discussion clearly in the video.

*Cognition EER to Incorporating Different Design Ideas and Building a Model*

The ultimate task in this sequence of experiments was to construct a model representing the greenhouse effect. During model construction, the main decisions to make concerned what materials to use, how to use each material, and for what reason each was used. When this group rebuilt their model on Day 4, they combined the instructed materials (a plastic fish tank, a black mat, and a white foam) and some other provided materials (tin foil, plastic wrap, and blue felt). They had come up with the idea about testing the thermal effect of a mirror and did test it in the tank, but they did not include the mirror when they rebuilt the model on Day 4.

On Day 5, while the lamp was turned off for the flasks and water in the flasks was cooling off, Grace asked the group; "So what object do we put in the tank...?" The group was still focused on waiting for the flasks to cool down, and there was not much response. After another six minutes, the temperatures started to level off, but the flasks had not cooled for 25 minutes yet. The group was still waiting. Grace then started to sketch her ideas on a piece of paper and showed it to other students in the group.

- Grace lays the paper in front of Sara and explains it to Sara.
- Sara: "Yeah. I like that."
- Grace then lays the paper in front of Jill and explains it to her.
- Jill looks at the paper and then asks: "Are we gonna use the mirror or not?"
- Grace pauses and then grabs back the paper and draws more on it.
- Ashley signs to Grace and gestures cutting. Interpreter speaks: "Just cut the foil."
- Jill talks more about the model but is inaudible.
- Sara: "If we wanted to use the mirror... (inaudible), then what if we put like ..." [gestures hands on the table]
- Grace listens and draws more on paper.
- Jill: "What if we put it like.... both the mirror and the foil…" [talks more but inaudible.]

In this EER, while the group was waiting for the flasks to cool down, Grace started collecting design ideas for their final model, directing the group to the next step of their activity. In these two days up to the point where they started to build their final model, the group had tested different materials and discussed different ideas to build their model. They also learned more about the materials throughout the experiments. Putting all the ideas together in a final model that represents the greenhouse effect, testing the model, and supporting their design with data, was their final challenge. They needed to complete it before the class period ended. Grace's first move was a jumpstart. When the discussion started, Jill, Sara, and Ashley made follow up moves (suggesting and coordinating different ideas) as shifts in direction, while Grace continued incorporating the group's ideas in her drawing. The explicit expressions were drawing, signing, verbal, handing out the sketch paper, accompanied by gesturing the shapes of the materials and the model. The effects were a sketch of their final model combining the ideas of everyone, and a model built based on it. This EER is illustrated in Figure 12.

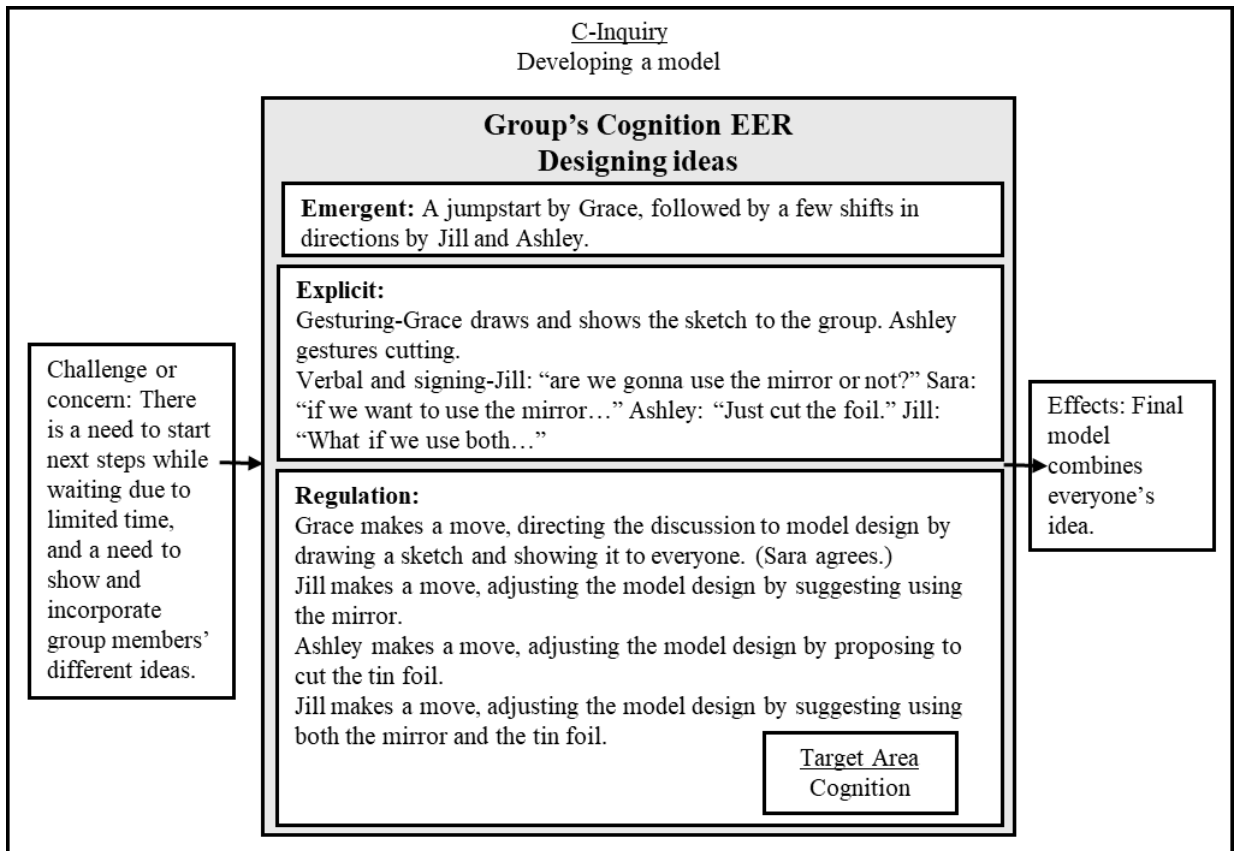

Figure 12. Group cognition EER in designing ideas.

Following this snippet, the group had more discussions about what materials to use and where and how each material should be placed in the model. Active discussions lasted throughout the rest of the experiment. The group's final model included both mirror and tin foil to represent different objects. See Figure 13.

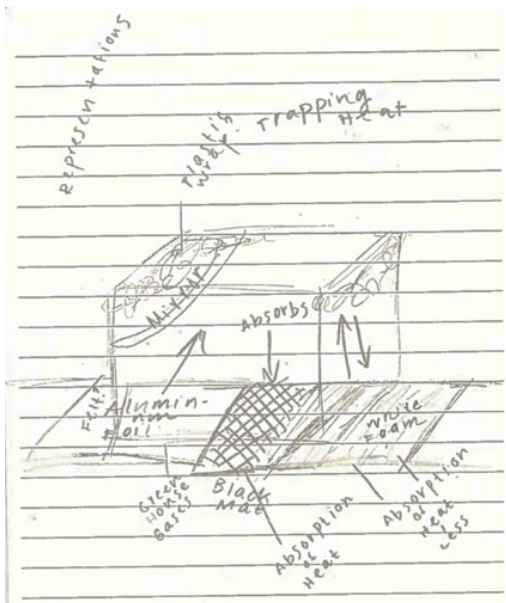

Figure 13. Grace's and Sara's journal showing the final model

As shown in Grace's drawing (left), they included a mirror in the top left corner, and a piece of aluminum foil at the left side of the bottom layers—the two materials Jill and Ashley brought up in the clip, and the group discussed how to incorporate them. The model also has a fish tank, a black mat, a white foam, a blue felt, and plastic wraps. Grace's annotations show their functions such as the black mat absorbs the heat and the white foam absorbs the heat less. Sara's journal (right) has textual descriptions of some materials about what they represent respectively. In contrast to Brittany's example that we presented earlier in this paper on the same task (building the group's final model as the end of the greenhouse effect model construction activity sequence), where Brittany controlled the process and built the model based on her idea and refused to explain her idea to the group, this group (Grace *et al.*) built their final model incorporating everyone's idea based on their discussion, reasoning, and data.

According to Grace's and Sara's journals in Figure 13, they used a piece of blue felt to represent water and a piece of aluminum foil on top of the felt to represent the source creating greenhouse gases. They made the claims based on their observation and comparison of the data. They tested two rounds of heating up and cooling down on the model, one round with the aluminum foil and one without. Their data showed that the model *with* the aluminum foil and the flask of water *with* the tablets both had a greater rate of change in temperature, compared to the ones without them (see Figure 10 right, bottom section). Therefore, they concluded that the tin foil represents the source of greenhouse gases (see Grace's and Sara's descriptions in Figure 13).

*Emotion*

The group's claim about the tin foil representing the source of greenhouse gases in the model was based on a phenomenological match in data. The representation was not conventional and had caught the instructor and the TA's attention. They visited the group and asked the group's explanation, but the group could not provide one.

We did not observe group members regulating each other's emotions, but we observed some expressions of their emotions. In the end of the class, on one hand, the group celebrated their success in that the data matched their proposed model of the greenhouse effect. Sara and Grace expressed their happiness with the results and their relief at the task being completed. They cheered each other up by giving high fives and saying, "I am so proud of us." On the other hand, they might have been aware that their explanation of the model representation was not strong. Jill expressed her feelings in her journal, but she did not explicitly show them during the activity. Jill wrote in her journal,

> The most frustrating part about this morning was that even after we finished our model, [instructor/TA] keeps coming around asking us questions such as why our model worked and what each … represented and why it represented that. It was frustrating because we were finished, and we knew our model worked. We didn't want to have to explain anything beyond that. I responded for these frustrations by answering the questions anyway even though I didn't want to! I could improve upon my experience and avoid frustration by being 100% confident about how everything I did fit together and what exactly I did and why. I should be able to know why something does or doesn't work. I can do this by analyzing and organizing my data better. To be honest, I am relieved that this climate change exercise is over, even though it was rewarding to finally get it right [Jill, Day 5].

## Discussion

Our case study demonstrated how the EER framework can be used to examine the regulatory moves made by group members toward achieving shared learning goals. We were able to capture the in-the-moment regulations that collectively drove the group towards completing a group task (constructing a satisfactory model), achieving shared goals (representing the greenhouse effect), and creating opportunities to learn (support model construction by data).

*Triggers and Foci of EER in Collaborative Inquiry*

In our case study, the challenges and concerns that triggered the typical EERs in the context of collaborative inquiry were related to the open-endedness of the experiment (e.g., free choice of materials, unknown experimental results) and collaboration (e.g., forming a new group, negotiating ideas from different people). Our case study group demonstrated, more or less, inquiry practices 0-8 during the two days (see Appendix B for examples), but the typical EERs had led the group to make adjustments in the two practices—developing a model and carrying out experiment—aligned with the tasks designed in the model activities: to build a model and test temperature data of different materials and different versions of the model. See a summary of the typical EERs in relation to collaborative inquiry in Figure 14.

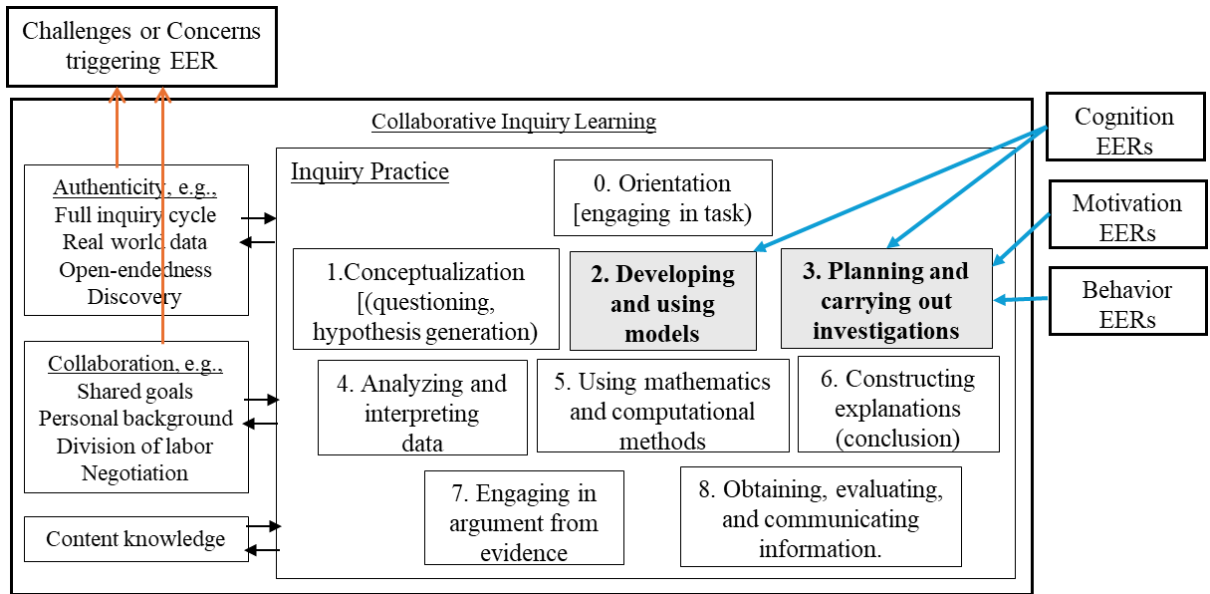

Figure 14. Typical EER in collaborative inquiry learning

Challenges and concerns were likely to arise in these two practices, and some on-the-spot adjustments were likely to be considered necessary and therefore being made. In contrast, challenges regarding other practices could be considered less critical, or beyond the scope of the activity, and adjustments were less likely to be made. For example, on Day 5, Ashley asked questions about "interpreting data" and "constructing explanations"—why the plain water heated up more quickly than the water with the tablets—but the group did not take up her questions to have a discussion nor make any adjustments.

Among the identified typical EERs, students focused on regulating their cognition when "developing a model." They focused on regulating their motivation, behavior, and behavior when "carrying out investigations." As a phenomenon, the group's typical social EERs were on Day 4 and their typical cognitive EERs were on Day 5; their typical motivation EERs were gentle and combined with a behavior EER, and the other typical behavior EER was at the beginning of Day 5.

The target areas and inquiry stages show the foci of the typical EERs. This case study demonstrates an approach to studying triggers and foci of socially shared regulative learning in collaborative inquiry (Vauras *et al.*, 2021). The EERs demonstrated that the group members actively engaged each other's ideas (Goos *et al.*, 2002), collectively strategized measurements and understanding the equipment's behavior (Van De Bogart *et al.*, 2017), and jointly negotiated, coordinated, and regulated their collaborative pursuit (Volet *et al.*, 2013).

*The Role of Social EERs*

The EER target area *social* has been demonstrated as an important aspect of emergent regulation in student groupwork in our data, both in the all-speaking student groups to mitigate a social conflict (Cao *et al.*, 2019), and in the mixed groups (this case study) to build social bonds between new group members. In this case study, the social EERs (i.e., asking names and majors, teaching and learning how to sign) were not part of inquiry practice, but they helped to keep the students in a collegial group during their collaborative inquiry. Indeed, in a previous quantitative study on student discourse group roles investigated in the same context, we found that students had more social discourse when new groups were formed (Wan *et al.*, 2025). However, whether the social EERs were intentional to build or improve the collaboration is uncertain (for example, they could just want to chat and kill the waiting time).

*Emergent Questions that Could Potentially Be Developed to an EER*

We noticed instances where students asked questions that could potentially be developed to an EER, but the group did not pursue that. For example, on Day 5, Ashley asked the group why the temperatures were higher in the plain water—to Sara: "What do you think?"; to the group: "But, why...?" "I mean maybe there's something else in there? I don't know. I don't know why." "So, this one is platonic. That one ... But why is ... Do we know yet?" The group did not pursue those questions. The emergent questions Ashley asked about data trends could direct the group to "interpreting data" and "constructing explanations" in inquiry practices, if the group pursued them. Those questions demonstrated the "authenticity" in collaborative inquiry—which involves real-world data, open-ended questions, and new discovery. Those questions could lead to fruitful discussions beyond completing the assigned tasks (i.e., building a model).

Jill also asked a couple of questions later in Day 5 about how the group thought about the greenhouse effect and climate change, connecting their experiments to the broader science concepts. Those questions were also fruitful and had the potential to be developed into an EER, directing the group to "obtaining, evaluating and communicating information" in collaborative inquiry, but the group did not follow up and discuss those questions.

*Mixed Methods of Communication*

The mixed communication methods in the group appeared to have better engaged the group members in social conversation. The group seemed more engaged in their social conversation by signing their name and learning how to sign. Ashley, being a member of the DHH community, also gave the group more social topics to talk about. This feature can potentially be applied to a context where group members are from different communities and/or speak different languages. Similar social interactions can be initiated from group members' diverse backgrounds and thus enhance the establishment of their collaboration.

We also speculate that some adjustments in this group (e.g., changing seats, sketching the model, writing down temperatures) could be due to their reliance on both verbal and non-verbal communications.

*Implications*

The moments where EER instances occur are when students take control and adjust the activity. Those moments often reflect the fact that the students have noticed something that bothered them and thus they take actions to make an adjustment to improve the situation. To promote more productive group work, instructional interventions can be designed around the moments when EER occurs. Strategic instructional questions can be asked to prompt, scaffold, or extend an EER. For example, when an adjust has been made due to an EER, the instructor could probe their thinking by asking "why did you change that?" In the videos, the IMPRESS instructor and TAs demonstrated many such kinds of interventions, which had led the students to articulate their thinking more explicitly. When a potentially productive EER was left incomplete, the instructor could provide scaffolding. In the example where Ashley asked questions about the temperature trends, the instructor could prompt the group to read the cold tablet package instruction to find out the chemical components of the tablets, and research about those components' thermal effects when dissolved in water. This kind of spontaneous scaffolding was observed less, perhaps due to the fact that the instructor or TA might have to come up with ideas on the spot, or the design of the activity was to not provide much scaffolding.

## Conclusion

The emergent explicit regulation framework provides a novel approach to analyze the momentary development of socially shared regulations in small groups. EER zooms in the moments where a regulatory move or multiple moves are developed from implicit to explicit (i.e., from facing a challenge to making adjustments) and directs the group to achieve their goals. Connecting regulating target areas with elements of collaborative scientific inquiry helps identify triggers and foci of students' regulation in this context. In our case study, Typical EER instances were identified in cognition, behavior, motivation, and social areas to handle the challenges in developing a model and carrying out investigation. The typical EERs were completed by multiple group members toward achieving shared group goals.

## Limitation and Future Work

The EER framework has been generated and tested in the IMPRESS program data with different groups at different times and in different activities. The case study in this paper was on one group in two activities in two consecutive days. The group members were all female. There was one signing student in the group and a professional interpreter. The case study conclusions are limited to this context.

This paper is intended to describe the developmental features of the typical EERs in detail. Quantitative coding of EER instances in a broader context will be valuable but beyond the scope of this study.

## Acknowledgements

We thank Rochester Institute of Technology and the IMPRESS program who supported our research study during all these years. We also thank the PEER professional development program who initially brought us together to analyze the videos. AAPT travel funds supported multiple presentations of this work. Other than the authors, Ruvim Popesku has participated in part of the early data analysis and contributed to the project.

# Appendices

Appendix A:

Descriptions of scientific inquiry practices 0-8, referencing NGSS and Pedaste *et al.*'s review paper (2015).

0. Students can be expected to form groups, read or listen to instructions, and engage in activity.

1. Students can be expected to ask scientific questions that the answers lie in explanations supported by empirical evidence, including evidence gathered by others or through investigation, and make hypotheses.

2. Students can be expected to evaluate and refine models through an iterative cycle of comparing their predictions with the real world and then adjusting them to gain insights into the phenomenon being modeled.

3. Students can be expected to design investigations that generate data to provide evidence to support claims they make about phenomena. In laboratory experiments, students are expected to decide which variables should be treated as results or outputs, which should be treated as inputs and intentionally varied from trial to trial, and which should be controlled, or kept the same across trials.

4. Students can be expected to expand their capabilities to use a range of tools for tabulation, graphical representation, visualization, and statistical analysis.

5. Students can be expected to use mathematics to represent physical variables and their relationships, and to make quantitative predictions. Computers and digital tools can enhance the power of mathematics by automating calculations, approximating solutions to problems that cannot be calculated precisely, and analyzing large data sets available to identify meaningful patterns.

6. Students can be expected to develop an explanation including a claim that relates how a variable or variables relate to another variable or a set of variables.

7. Students can be expected to engage in argumentation, a process for reaching agreements about explanations and design solutions.

8. Students can be expected to read, interpret, and produce scientific and technical texts, and to communicate clearly and persuasively.

Appendix B

Examples of the group's inquiry practices during Day 4 and Day 5. Evidence taken from the videos and their journals.

| Inquiry practice | Day 4 | Day 5 |
|---|---|---|
| 0 Orientation | Come in, listen to the instructor announcing tasks of the day: choose to work with someone they haven't, sit at the table, instructed task. | Come in, listen to the instructor announcing tasks for the day. |
| 1 Conceptualization (Questioning and hypothesizing) | Check and understand the task and questions. | Check and understand the task and questions, ask questions and speculate reasons the plain water is warmer, ask questions about greenhouse gases and climate change. |
| 2 Developing and using models | Discuss model building ideas, use the tank, tin foil, baking paper, wrapped with plastic wraps to build the model, test the model with and without the plastic wraps. | Discuss model building ideas, use the tank, black mat, white foam, blue felt, tin foil, mirror, plastic wrap to build the model, test temperatures of the model with and without the tin foil. |
| 3 Planning and carrying out investigations | Set up equipment, turn on and off the heating lamp, record data. Test the white foam, the flip side of the black mat, then a mirror at the bottom of the tank. After they build the model, also test the in the model. | Set up equipment, turn on and off the heating lamp, record data. Test the two flasks with and without the cold tablets (release $CO_2$). After they build the model, test the model with and without the tin foil. |
| 4 Analyzing and interpreting data | Compare heat-up and cool-down rates of different materials (white foam, mirror, tin, felt, plastic wrap) in the model. | Compare heat-up and cool-down rates of water in flasks with and without $CO_2$, and the model with and without the tin foil. |
| 5 Using mathematics and computational thinking | Consider and compare the rates of change in temperature when testing different materials, use computer programs to collect temperature data. | Consider and compare the rates of change in temperature as a function of time when testing the water and the model, use computer programs to collect temperature data. |
| 6 Constructing explanations | Discuss roles of materials in representing the Earth atmosphere, for example, a mirror represents cloud to reflect heat back, plastic wrap is to trap the heat. | Speculate something between the flask and water that makes the plain water heated up faster, discuss roles of materials in representing the Earth atmosphere, for example, the tin foil represents the source of greenhouse gases. |
| 7 Engaging in argument from evidence | Argue gently about why they want to test the flip side of the black mat when they remove the white foam. | Argue about what materials to use in the model, and where to put them, for example, whether to use the mirror, where to put the mirror, where to put the tin foil, and how much area the foil should cover. |
| 8 Obtaining, evaluating, and communicating information | Discuss model ideas. Write in journals about their experiment. | Find the boiling point of water from the internet, sketch models to communicate model design ideas, write journals about their experiment. |